# Enhancing Celestial Imaging: High Dynamic Range with Neuromorphic Cameras

Authors:
Satyapreet Singh Yadav[1], Nirupam Roy[2], Chetan Singh Thakur[1]

[1]Department of Electronic Systems Engineering, Indian Institute of Science

[2]Department of Physics, Indian Institute of Science

Conventional frame-based cameras often struggle with limited dynamic range, leading to saturation and loss of detail when capturing scenes with significant brightness variations. Neuromorphic cameras, inspired by human retina, offer a solution by providing an inherently high dynamic range. This capability enables them to capture both bright and faint celestial objects without saturation effects, preserving details across a wide range of luminosities. This paper investigates the application of neuromorphic imaging technology for capturing celestial bodies across a wide range of flux levels. Its advantages are demonstrated through examples such as the bright planet Saturn with its faint moons and the bright star Sirius A alongside its faint companion, Sirius B.

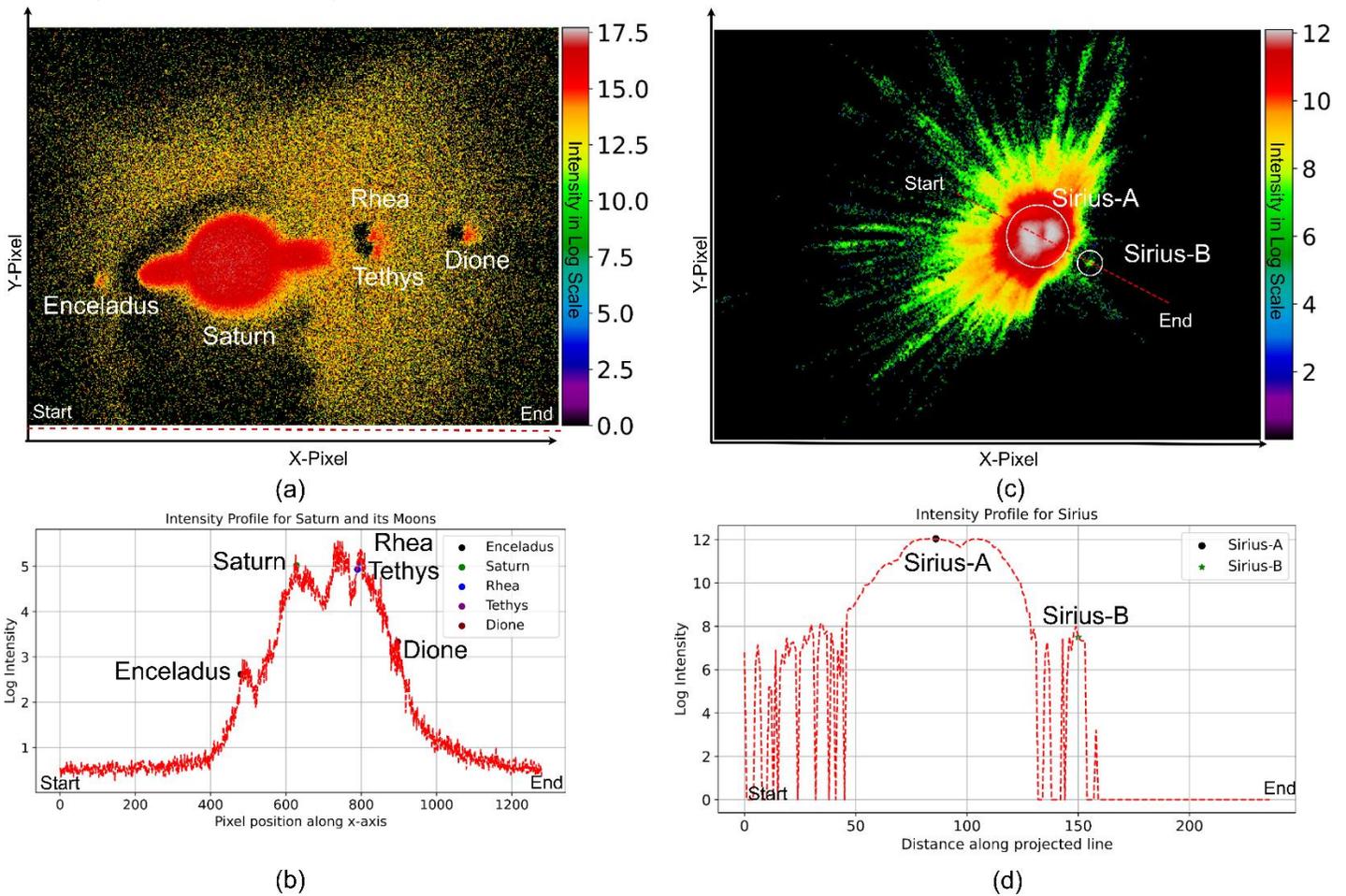

Figure 1. High Dynamic Range (HDR) Imaging Using a Neuromorphic Camera | The neuromorphic camera was interfaced with the 1.3m DFOT telescope at ARIES, India, with a plate scale of 0.195". **(a)** A 1.46' x 1.95' field-of-view (FoV) of log-transformed image of Saturn and its moons—Enceladus, Rhea, Tethys, and Dione —with stellar magnitudes of 1.16, 11.94, 10.13, 10.58, and 10.79, respectively. The data was captured on 2nd June 2024 at 21:30:36 UTC observed over a 5-second duration. **(b)** The intensity plots for Saturn and its moons along the x-axis show the distinct intensity profiles of the objects, where Saturn's profile dominates due to its higher brightness, while the moons exhibit varying intensity levels consistent with their respective magnitudes. **(c)** A 1.46' x 1.95' FoV of log-transformed image of the Sirius binary star system, comprising Sirius A with a stellar magnitude of -1.46 and Sirius B with a magnitude of 8.41 in the B band. The data was captured on 28th November 2023 at 18:10:34 UTC over a 9.5-second duration. **(d)** The intensity plots for Sirius A and Sirius B along a line passing through the center of the binary star system demonstrate the camera's ability to resolve high-contrast objects in the same frame. Sirius A, being significantly brighter than Sirius B, shows a wider and more pronounced intensity profile. The sharp and narrow profile of Sirius B highlights the camera's ability to resolve faint sources next to extremely bright objects. The neuromorphic camera successfully captured objects of varying brightness without saturation, achieving a dynamic range (DR) exceeding 86.24dB.

Charge-Coupled Device (CCD) cameras have revolutionized the field of astronomy since their invention, enabling significant breakthroughs and discoveries [1-2]. While CCDs offer high quantum efficiency, they are limited by issues such as saturation effects, bleeding artifacts [3], fixed frame rates, and a limited dynamic range of ~60 dB. The dynamic range of a camera defines its ability to simultaneously capture information from bright and faint sources in a single frame without saturation.

In imaging scenarios with both very bright and faint objects, long exposure times often cause bright objects to saturate, reducing the visibility of faint sources. This issue is further worsened by bleeding effects, where excess charge from saturated pixels spreads into adjacent regions. A common approach to address this challenge is capturing a series of images with varying exposure times and stacking them [4]. This method enables the detection of both bright and faint objects within the scene using High Dynamic Range (HDR) signal processing techniques, shifting the complexity from sensor to the post-processing stage.

The human eye, by contrast, possesses an exceptionally high Dynamic Range (DR) of >120 dB [5] as it perceives brightness logarithmically. Inspired by the capabilities of the human eye, the neuromorphic community developed the neuromorphic camera. This camera converts incoming light intensity into a logarithmic voltage, effectively preventing saturation. It is designed to detect contrast changes and generates output in the form of spikes or events. A positive or negative event is triggered when the contrast at a pixel increases or decreases beyond a preset threshold in the neuromorphic camera [6].

Each active pixel is read asynchronously using the asynchronous event representation (AER) protocol, ensuring that only pixels generating events are transmitted to the AER bus, resulting in a low and efficient data rate. This asynchronous readout provides the camera with an exceptionally high temporal resolution in the microsecond (µs) range, while the logarithmic conversion ensures a high DR [7]. With total power consumption of just a few milliwatts, the neuromorphic camera's unique characteristics make it particularly well-suited for capturing dynamically changing scenes across a wide range of illuminations. Its small form factor and low power requirements offer significant advantages over conventional frame-based cameras, particularly for space-based astronomical tasks.

In this work, we explore the use of a neuromorphic camera for HDR imaging of celestial objects in astronomical applications through ground-based observatories. The camera was integrated at the focal plane of the 1.3m Devasthal Fast Optical Telescope (DFOT), located at the AryaBhatta Research Institute of Observational Sciences (ARIES) in Uttarakhand, India. The telescope's focal ratio is 4, making it a fast optical system, and it provides a pointing accuracy of 10" rms and tracking accuracy of 0.5" rms [8]. For imaging, B and V band filters were used. The observations were performed on the nights of 28th November 2023 and 2nd June 2024 using a neuromorphic camera consisting of 1280 x 720 pixels, with a pixel pitch width of 4.8 µm. With the neuromorphic camera, the setup achieved a plate scale of 0.195", capturing a FoV of 4.15' x 2.33'.

The telescope tracked the celestial bodies under observation on both nights. The binary star system Sirius comprises the bright primary star Sirius A, with a magnitude of -1.46, and its faint companion, the dwarf star Sirius B, with a magnitude of 8.41 in the B band. Observations were carried out on 28th November 2023 at 18:10:34 UTC (Coordinated Universal Time) over a duration of 9.5 seconds, with both On-contrast and Off-contrast thresholds set to 150.

Saturn with magnitude 1.16 and its moons—Enceladus with magnitude 11.94, Rhea with magnitude 10.13, Tethys with magnitude 10.58, and Dione with magnitude 10.79—were observed in the V band for 5 seconds on 2nd June 2024 at 21:30:36 UTC. As the celestial bodies imaged by the neuromorphic camera jittered slightly within the FoV due to atmospheric tilt and tip, the camera generated both positive and negative events. The On-contrast threshold was set to 50, while the Off-contrast threshold was set to 70. These event data were converted into images by accumulating the events on each pixel as they occurred, resulting in a grayscale representation of the scene.

To enhance visual clarity and highlight scale variations, false color coding was applied to the images.

DR of a camera is typically defined as ratio of maximum intensity ($Imax$) that the camera can capture to the Root Mean Square (RMS) noise level. It is expressed in decibels as

$$DR\ (dB) = 20\ log10\left(\frac{Imax}{RMS\ noise}\right) \rightarrow (1)$$

The difference in apparent magnitudes ($m_1$ - $m_2$) of two celestial bodies is defined in terms of the ratio of their flux values ($I_1$ and $I_2$) as:

$$m1 - m2 = -2.5\ log10\left(\frac{I1}{I2}\right) \rightarrow (2)$$

Figure 1 demonstrates the camera's high dynamic range (HDR) capabilities through observations of Saturn and the Sirius binary system. In Figure 1(a), the flux ratio between Saturn (the brightest object) and Enceladus (the faintest) is approximately 20,500 (from Eq.2), which corresponds to a dynamic range exceeding 86.24 dB (from Eq.1). This is evident in Figure 1(b), where Enceladus' intensity is clearly above the noise floor, despite poor seeing conditions greater than 2". The intensity profile for Sirius (Figure 1(d)) also reveals a wider range of intensities further showcasing the system's HDR. Even with a bright full moon background during the Sirius observation, the camera clearly resolves Sirius B, demonstrating its ability to capture distinct features within a complex scene containing a wide range of intensities.

Capturing celestial bodies with a neuromorphic camera produces images with a high dynamic range, unaffected by saturation at the camera level. However, during ground-based observations, the DR is limited by factors such as atmospheric seeing conditions, sensor noise, and pixel response variations. Despite poor seeing conditions (>2") on June 2, 2024, and full moon night on 28th November 2023, the camera delivered excellent results. The neuromorphic camera's HDR capability, combined with its low power consumption, high temporal resolution, and reduced data rate, makes it an ideal sensor for space applications.

**Acknowledgments** | We sincerely thank Dr. Bikram Pradhan from the Indian Space Research Organization and Dr. T.S. Kumar from ARIES for providing us the opportunity to visit ARIES over the span of two years to conduct neuromorphic imaging of celestial objects.

**REFERENCES**

1. Mackay, C.D.: Charge-coupled devices in astronomy. Annual review of astronomy and astrophysics 24(1), 255-283 (1986).
2. Catterall, A.: Ccd imaging from the city. In: The Art and Science of CCD Astronomy, pp. 101–110. Springer, New York (1997).
3. Krishnamurthy, A., Villasenor, J., Seager, S., Ricker, G., Vanderspek, R.: Precision characterization of the tess ccd detectors: Quantum efficiency, charge blooming and undershoot effects. Acta Astronautica 160, 46–55 (2019).
4. Vítek, S., & Páta, P. (2016). Realization of High Dynamic Range Imaging in the GLORIA Network and Its Effect on Astronomical Measurement. *Advances in Astronomy*, *2016*(1), 8645650.
5. Darmont, A. (2013). High dynamic range imaging: sensors and architectures.
6. Gallego, G., Delbrück, T., Orchard, G., Bartolozzi, C., Taba, B., Censi, A., Leutenegger, S., Davison, A.J., Conradt, J., Daniilidis, K., et al.: Event-based vision: A survey. IEEE transactions on pattern analysis and machine intelligence 44(1), 154–180 (2020).
7. Finateu, T., Niwa, A., Matolin, D., Tsuchimoto, K., Mascheroni, A., Reynaud, E., & Posch, C. (2020, February). 5.10 A 1280× 720 back-illuminated stacked temporal contrast event-based vision sensor with 4.86 μm pixels, 1.066 GEPS readout, programmable event-rate controller and compressive data-formatting pipeline. In 2020 IEEE International Solid-State Circuits Conference-(ISSCC) (pp. 112-114). IEEE.
8. Joshi, Y., Bangia, T., Jaiswar, M., Pant, J., Reddy, K., Yadav, S.: Aries 130-cm devasthal fast optical telescope—operation and outcome. Journal of Astronomical Instrumentation 11(04), 2240004 (2022).